\documentclass[letterpaper]{JHEP3}

\newcommand{\beq}{\begin{equation}}
\newcommand{\eeq}{\end{equation}}
\newcommand{\cR}{{\cal R}}
\usepackage{graphicx}
\usepackage{amsmath}
\usepackage{amsfonts}
\usepackage{epsfig}

\title{de Sitter gravity from lattice gauge theory}

\author{Simon Catterall \\
Department of Physics, Syracuse University, Syracuse, NY13244, USA \\
E-mail: \email{smc@phy.syr.edu}
}
\author{Daniel Ferrante \\
Department of Physics, Syracuse University, Syracuse NY13244, USA\\
E-mail: \email{ddferran@syr.edu}
}
\author{Arwen Nicholson \\
Department of Physics and Astronomy, Edinburgh University, Edinburgh EH9 3JZ,
UK\\
E-mail: \email{arwen.e.nicholson@gmail.com}
}

\abstract{We investigate a lattice model for Euclidean quantum gravity based
on discretization of the Palatini formulation of General Relativity.
Using Monte Carlo
simulation we show that while a naive approach fails to lead to
a vacuum state consistent with the emergence of classical spacetime,
this problem may be evaded if the lattice action is supplemented by
an appropriate counter term. In this new model 
we find regions of the parameter space which admit
a ground state which can be interpreted as 
(Euclidean) de Sitter space.
}

\begin{document}

%
\section{Introduction}

The problem of constructing a theory of gravity which is consistent
with quantum mechanics has a long history. The non-renormalizability
of quantum general relativity can be traced to the fact that
the Newton constant carries the dimensions of length squared
corresponding to an action linear in the curvature. Actions containing
higher powers of the curvature are usually thought to lead to
violations of unitarity and can at best be thought of
only as effective field theories \cite{Donoghue:1994dn}.

Alternative approaches to gravity have tried to highlight the
similarities to gauge theory, where in the case of gravity, the local
symmetry corresponds to Lorentz invariance. In this case the
corresponding 
gauge field is called a spin connection. It is necessary
to introduce such an object when discussing fermions
in curved spacetime where it is partnered by a new field the
so-called vierbein which describes the local Lorentz frame and
from which the usual metric tensor can be reconstructed.
An action coupling the vierbein to the Yang-Mills curvature associated
with this spin connection can be written down and shown
to reproduce the usual Einstein equations under certain conditions --
that the vierbein, regarded as a matrix, be invertible and
that the torsion -- the antisymmetrized covariant derivative of the vierbein, be
zero. It is a first order formulation as both the vierbein and
spin connection are to be varied independently to determine
the classical equations of motion.

This recasting of gravity in the language of gauge theory
is termed the Palatini or Palatini-tetrad formulation
of General Relativity \cite{tetrad}. In principle this
approach gives a natural starting point for a non-perturbative
study of quantum gravity since it is possible to discretize the
theory while maintaining exact gauge invariance using techniques
similar to those employed for (lattice) QCD. As such it is complementary
to lattice approaches based on Euclidean dynamical triangulations
see eg. \cite{Catterall:1994pg,Ambjorn:1991pq,deBakker:1994zf,Agishtein:1992xx} and causal 
dynamical triangulations 
\cite{Ambjorn:2008wc,Ambjorn:2008sw,Ambjorn:2007jv} and references therein.

A number of proposals have been made earlier along these lines
\cite{Smolin:1978kq,Menotti:1986bm,Menotti:1986uc,
Menotti:1985yb,Caracciolo:1987nj}
and numerical studies performed \cite{Caracciolo:1987ng,Caracciolo:1987qi,
Caracciolo:1988dg}.
Unfortunately this previous
work encountered several difficulties; chief among
these was the observation that the expectation value of the
vierbein vanished in the classical weak coupling limit. This condition
by itself is sufficient to invalidate the connection between the gauge theory
and gravity as we will discuss in the
next section. 

In this work we have explored the phase structure of
a lattice model in which the discrete Palatini action is supplemented by
an additional Yang-Mills term. We find a region of the
parameter space of the model where the vierbein is non-zero, the
torsion small and the curvature constant. 

The paper starts with a review of the Palatini formalism in the continuum,
and describes its discretization on the lattice. The non-perturbative
structure of this theory is then explored using Monte Carlo simulation
and we show the classical ground state of the theory corresponds to
vanishing vierbein and a  curvature which approaches its kinematic
limits \footnote{In Euclidean space we replace
the Lorentz group $SO(3,1)$ by its compact analog $SO(4)$. 
In the context of the
lattice theory this ensures that the local curvature is bounded
both above and below}. Thus the resulting lattice theory does not possess an 
appropriate ground
state which can be identified with a continuum
geometry. We then modify the model to incorporate
an additional curvature squared operator and survey the expanded
phase diagram finding indications of a region where the ground state of
theory can be identified with Euclidean de Sitter space. In the final
section of the paper we discuss what these results imply and what
further work must be done to solidify any possible connection to
a theory of gravity.

\section{Review of the Palatini formalism} 

The Palatini action can be regarded as the most general action
constructed from the curvature of
the spin connection and the vierbein which
is invariant under local Lorentz transformations and 
whose definition is independent of the choice of
background metric. It takes the form

\begin{equation}
  S_{\rm{Palatini}} = \frac{1}{l^2_p}\, \int d^4x\,
    \epsilon_{\mu\nu\lambda\rho}\, \epsilon^{ijkl}\, e^i_{\mu}\, e^j_{\nu}\,
    \biggl(R^{kl}_{\lambda\rho} - \frac{\Lambda}{6}\, e^k_{\lambda}\,
    e^l_{\rho}\biggr)
    \label{tet} 
\end{equation}
with $l_p=\sqrt{4\pi G}$ the Planck length.
The usual metric of General Relativity is then related to the vierbein via
\begin{equation}
  g_{\mu\nu} = \eta_{i\, j}\, e^i_{\mu}\, e^j_{\nu} 
\end{equation}
where $\eta_{i\, j}=\delta_{ij}$ is a flat (Euclidean) metric which henceforth
we shall leave implicit in formulae.
This relation is clearly invariant under local
$SO(4)$ gauge transformations mediated by the spin connection $\omega_{\mu}$
where 
\beq
\omega_{\mu} =
\sum_{i<j} \omega_{\mu}^{ij} T^{ij}\qquad i<j=1\ldots 4\eeq
with $T^{ij}$ being appropriate generators of
$SO(4)$. The curvature $R_{\mu\nu}^{ij}(\omega)$ appearing in eqn.~\ref{tet} 
is the
usual Yang-Mills field strength associated with the spin connection.
\begin{equation}
  R^{ij}_{\mu\nu} = \partial_{\mu} \omega_{\nu}^{ij} - \partial_{\nu}
    \omega_{\mu}^{ij} + \big[\omega_{\mu}^{ik} , \omega_{\nu}^{kj}\big]
\end{equation}
Variation of this action with respect to the spin connection yields
the torsion free constraint $D_{\left[\mu\right.} e^i_{\left.\nu\right]} = 0$
which can be used to solve for the spin connection in terms of the
vierbein {\it provided} the inverse $e^\mu_i$ exists. 
The solution is
\begin{equation}
  \omega_{\mu\, ij} (e) = \frac{1}{2}\, e^{\nu}_ie^{\rho}_j\,
    \big(\Omega_{\mu\nu\rho} - \Omega_{\nu\rho\mu} + \Omega_{\rho\mu\nu}\big) \; ;
\end{equation}
where,
\begin{equation}
  \Omega_{\mu\nu\rho} = \big(\partial_{\mu} e_{\nu}^k - \partial_{\nu}
    e_{\mu}^k\big)\, e_{\rho\, k} \; 
\end{equation}
Variation of the action with respect to the vierbein
then yields the equation,
\begin{equation}
  \epsilon^{ijkl}\, \epsilon_{\mu\nu\lambda\rho}\, \biggl(R^{ij}_{\mu\nu} -
    \frac{\Lambda}{3}\, e^i_{\mu}\, e^j_{\nu}\biggr)\, e^k_{\lambda} = 0
\end{equation}
which, in conjunction with the torsion free solution
given above $\omega = \omega_{\mu}(e)$,
just reproduces the usual Einstein equation for pure gravity with
a cosmological constant. The vacuum
solutions are then constant curvature spaces --
here Euclidean de Sitter space - the sphere $S^4$.

As MacDowell and Mansouri noted \cite{MacDowell:1977jt}
it is possible to forge an even stronger connection to Yang-Mills theory
by combining the $SO(4)$ spin connection $\omega_\mu$ and
vierbein $e_\mu$ into a single gauge field $A_\mu$ associated with
a enlarged
$SO(5)$ gauge symmetry 
\begin{equation}
A_\mu = \omega_\mu^{ij} T^{ij}+\frac{1}{l}e_\mu^{i}T^{5i}\qquad i, j = 1,\ldots 4 
\end{equation}
where $l$ is some scale
inserted to render the vierbein dimensionless.
In a similar fashion the $SO(5)$ curvature $F_{\mu\nu}$ can be decomposed
into components transforming under the same  $SO(4)$ subgroup
\begin{eqnarray}
F_{\mu\nu}&=&\left(R_{\mu\nu}-\frac{1}{l^2}e_{\left[\mu\right.}e_{\left.\nu\right]}\right)^{ij}
T^{ij}\nonumber\\
&+&\frac{1}{l}D_{\left[\mu\right.}e_{\left.\nu\right]}^{i}T^{5i}\qquad
i,j=1\ldots 4
\label{so5}
\end{eqnarray}
with $R$ the $SO(4)$ curvature.

An $SO(4)$ invariant action quadratic in this $SO(5)$ curvature
may then be written down
\begin{equation}
  S=\kappa\int\,d^4x \epsilon_{\mu\nu\lambda\rho}\, \epsilon^{ijkl5}
    F^{ij}_{\mu\nu}F^{kl}_{\lambda\rho}
    \label{MM}
\end{equation}
By expanding the $SO(5)$ field strengths according to
eqn.~\ref{so5} this
is easily seen to be nothing more than the 
Palatini action given in
eqn.~\eqref{tet} with the dimensionless coupling
$\kappa=\frac{1}{2}\left(\frac{l}{l_p}\right)^2$ giving the
size of the Universe in units of the Planck length
and $\Lambda=\frac{3}{l^2}$\footnote{ 
A term quadratic in the $SO(4)$ curvatures also appears 
$\int\epsilon_{\mu\nu\lambda\rho}\epsilon^{ijkl} 
R_{\mu\nu}^{ij}R_{\lambda\rho}^{kl}$ but can be neglected
as it corresponds to a topological
invariant - the Euler
number}
Notice that the classical equations of motion derived from
the Palatini action just correspond to
setting the $SO(5)$ curvature to zero.
 
This action while formulated in terms of an $SO(5)$ connection exhibits
an explicit breaking to $SO(4)$. It is natural to generalize this
action slightly to try and realize this
as a spontaneous breaking. The following $SO(5)$ action accomplishes this
at the expense of introducing an additional scalar field $\phi$.
\begin{equation}
  S=\int d^4x \epsilon_{\mu\nu\lambda\rho}\, \epsilon^{ijklm}
    F^{ij}_{\mu\nu}F^{kl}_{\lambda\rho}\phi^m
    \label{MM2}
\end{equation}
On the assumption that the scalar acquires a vacuum expectation
value of the form
$\phi^m=\kappa\delta^{5m}$ we recover the previous action. 

The appearance of a scalar field $\phi$ and an associated $SO(5)$
gauge symmetry seems at first sight to be somewhat arbitrary. However,
it is possible to show that this $SO(5)$ invariant theory arises
naturally as a compactification of a Chern-Simons theory in five dimensions.
This Chern-Simons theory is a generalization to five dimensions of Witten's
formulation of three dimensional gravity as a Chern Simons theory
\cite{Witten:1988hc, Chamseddine:1989nu}. In the Euclidean case 
considered here this amounts to constructing a topological gravity
theory in five dimensions with internal symmetry group $SO(6)$ and
action
\begin{equation}
  S_{\rm CS} = \int_{M_5} < \Omega\wedge \cR \wedge \cR  + \frac{3}{2}\,
    \Omega\wedge \Omega\wedge \Omega\wedge \cR + \frac{3}{5}\, \Omega\wedge
    \Omega\wedge  \Omega\wedge
    \Omega\wedge \Omega> 
\end{equation}
where for brevity we used the language of differential forms to
represent the contraction of spacetime indices with the
five dimensional epsilon symbol.
The Lie algebra valued connection is
\begin{equation}
  \Omega_{\mu} = \sum_{A<B} \Omega_{\mu}^{AB}\, J_{AB} \;;\; A, B = 1,\ldots, 6 \; ;
\end{equation}
with $J_{AB} = [\gamma_A , \gamma_B]$ the generators of $SO(6)$
and $\cR$ the curvature. The 
the angular brackets indicate that the group
indices are contracted with the  $SO(6)$ invariant tensor
$\epsilon_{ABCDEF}$. This Chern-Simons Lagrangian is related to the Euler
density in six dimensions via the relation,
\begin{equation}
  dL_{\rm CS} = \epsilon_{ABCDEF}\, \cR^{AB}\wedge \cR^{CD}\wedge \cR^{EF}
\end{equation}
Stokes' theorem then guarantees that the Chern-Simon's theory is
invariant under local gauge transformations up to possible
boundary terms.

It is useful at this point to decompose the connection in terms of
quantities transforming simply under an $SO(5)$ subgroup
\begin{equation}
  \Omega=\sum_{a<b}A^{ab}J_{ab}+\sum_a E^aJ_{6a}\quad a,b=1\ldots 5
  \end{equation}
  Similarly the $SO(6)$ field strength decomposes according to
\begin{equation}
  \cR = \sum_{a<b}(F^{ab} + E^{a} \wedge E^{b})J_{ab}+\sum_a DE^a J_{6a} 
\end{equation}
where $F$ is the $SO(5)$ curvature and $DE$ the torsion.
The action up to boundary terms becomes
\begin{equation}
  S_{\rm CS} = \int_{M_5} \epsilon_{abcde}\, \left(E^a \wedge F^{bc}\wedge
    F^{de} + \frac{2}{3}\, E^a\wedge E^b\wedge E^c\wedge F^{de} +
    \frac{1}{5}\, E^a\wedge E^b\wedge E^c\wedge E^d\wedge E^e\right) 
    \label{CS5}
\end{equation}

In this form it is clear that the action does indeed describe
a gravitational theory containing a
``Gauss-Bonnet form'', an Einstein term and a cosmological constant in five
dimensions. Furthermore, this theory is invariant under the 
full (Euclidean) de Sitter group not just local Lorentz transformations.

To make contact with the MacDowell-Mansouri form of the
Palatini action we must compactify this theory down to four dimensions.
Let us assume that the
five dimensional manifold is of the form $M_5 = M_4\times
S^1/\mathbb{Z}_2$ with $l_5$ the extent of the fifth dimension.
On the four dimensional boundaries we will assume that the gauge field
satisfies the following boundary conditions,
\begin{subequations}
  \label{bcs}
  \begin{align}
    \Omega_{\mu} &= \gamma_6\, \Omega_{\mu}\, \gamma_6 \;;\; \mu = 1,\dotsc, 4 \\
    \Omega_5 &= -\gamma_6\, \Omega_5\, \gamma_6
  \end{align}
\end{subequations}
This breaks the $SO(6)$ gauge symmetry down to
$SO(5)$ at the boundaries with
the only surviving four dimensional fields being 
\begin{equation}
  E_5^a \;,\;\; A_{\mu}^{ab} \; .
\end{equation}
Returning to the action in eqn.~\ref{CS5} it shoul be
clear that only the ``Gauss-Bonnet'' term survives at
tree level on the four dimensional boundaries. 
Indeed, this term becomes
\begin{equation}
  \int d^4 x\; \epsilon_{abcde}\, \epsilon_{\mu\nu\lambda\rho}\, E_5^a\,
    F^{bc}_{\mu\nu}\, F^{de}_{\lambda\rho}
\end{equation}
This is nothing more than the previous action with the fifth
component of the five dimensional vielbein playing the role of the
scalar field in the four dimensional
theory.
The development of a vacuum expectation value for
the scalar field could then be realized in terms of the
appearance of a non vanishing
Polyakov line in the fifth direction extending
between the two four dimensional
boundaries.
 
Thus the SO(5) invariant theory given in eqn.~\ref{MM2} can
be realized by an appropriate compactification of a topological gravity theory
in five dimensions. The possible existence of an underlying exact $SO(5)$
gauge symmetry in the four dimensional
theory has some advantages; it allows us to restrict
possible counter terms to those invariant under $SO(5)$, it makes the
choice of the Haar measure for $SO(5)$ natural in the path integral
and also, as we now explain, it makes the restoration of diffeomorphism
invariance a more likely possibility.

In \cite{Witten:1988hc} Witten showed that in fact diffeomorphism invariance
is intimately connected to invariance under both lorentz transformations
and local translations. 
In the four dimensional model considered here the result of a general 
coordinate transformation with parameter $-\xi^\nu$
acting on the spin connection and vierbein can be written
compactly
as a gauge transformation with parameter $\xi^\mu A_\mu$ 
acting on the $SO(5)$ gauge field $A=\omega+e$ plus a term
which vanishes on flat connections with $F=0$
\beq
\delta^\xi A_\mu =-D_\mu (\xi^\nu A_\nu)-\xi^\nu F_{\mu\nu}\eeq
Thus provided the theory is $SO(5)$ invariant and we consider only
small fluctuations around a flat background the theory will
automatically be invariant under general coordinate
transformations.

\section{Lattice theory}

It is straightforward to formulate
the Palatini action in the form given by
eqn.~\ref{MM} on a hypercubic lattice
as was first suggested in \cite{Smolin:1978kq}.
\beq
S_P=\kappa\sum_x\sum_{\mu\nu\rho\lambda}\epsilon_{\mu\nu\lambda\rho}{\rm Tr}\left(\gamma_5
U_{\mu\nu}U_{\lambda\rho}\right)\eeq
where 
\beq
U_{\mu\nu}=U_\mu(x)U_\nu(x+\mu)U^\dagger_\mu(x+\nu)U^\dagger_\nu(x)
\eeq
is a Wilson plaquette variable and takes its
values in the group $SO(5)$\footnote{For simplicity in this
study we have not symmetrized
our action to ensure rotational invariance nor have we
concerned ourselves at this point with issues of reflection positivity}. 
Notice that in our study we have
taken as generators of $SO(5)$ the matrices 
$T^{ij}=\frac{1}{4}[\gamma^i,\gamma^j]$
where $\gamma_i, i=1\ldots 5$ are the usual
four dimensional Dirac matrices together with the chiral matrix $\gamma_5$.
This differs from the earlier numerical work reported in 
\cite{Caracciolo:1987qi,Caracciolo:1988dg}
which utilized the vector representation of $SO(5)$
and implies that we are actually
simulating a lattice theory based on the {\it covering group} spin(5). 
As we will see this will turn out to be rather important.

Notice though that the introduction of a lattice has necessarily
broken the coordinate invariance of the theory in the base space.
Thus one should expect that quantum corrections will generate additional
operators whose structure depends on the existence of this background
lattice. Assuming that these operators depend only on the $SO(5)$
curvature there are 3 such terms in addition to
the Palatini action which are (marginally) relevant by
power counting; 
\begin{center}
\begin{itemize}
\item $\int d^4 x \epsilon_{\mu\nu\lambda\rho}{\rm
Tr}\left(F_{\mu\nu}F_{\lambda\rho}\right)$
\item $\int d^4 x {\rm Tr}\left(F_{\mu\nu}F_{\mu\nu}\right)$
\item $\int d^4 x {\rm Tr}\left(\gamma_5 F_{\mu\nu}F_{\mu\nu}\right)$
\end{itemize}\end{center}
The first of these is a topological invariant in the
continuum analogous to the
instanton number and hence can be
neglected. In principle we should include lattice versions of both
the second and third terms. In practice we have focused these
initial investigations on just the second term.  On the lattice
we implement it with the usual Wilson plaquette action
\beq
S_W= \sum_x\sum_{\mu<\nu} {\rm Tr} \left(2-U_{\mu\nu}-U^\dagger_{\mu\nu}\right)
\eeq
Our choice of this term has clear motivation; it clearly will favor
configurations in the large $\kappa$ limit with vanishing $SO(5)$
curvature -- a feature which will see shortly is
{\it not} true of the pure Palatini action. Such configurations
are minimally required to achieve a connection to classical gravity. 
It also has the merit of removing potential lattice
doubler modes which will be present in the Palatini action
as was first observed in \cite{Menotti:1985yb}.

Of course the presence of a Wilson term is not
compatible with coordinate invariance in the base space. It will
clearly be very important to test for a restoration of this
property in any continuum limit. Nevertheless the important point
to realize is that this term is {\it necessarily} induced in the lattice
theory and to have a hope to obtaining the correct continuum limit
we should include it in the bare lattice action and tune its coupling
appropriately as the lattice spacing is reduced.
The lattice action we consider then has the form
\beq
S=\kappa\left(\alpha S_W+(1-\alpha)S_P\right)\eeq
which allows us to interpolate between the pure Palatini action
defined by $\alpha=0$ and pure Wilson action when $\alpha=1$.

Finally, we need to give a presecription for extracting the various
$SO(4)$ components of the connection and curvature from the
basic $SO(5)$ variables in the theory. We used the simple
expressions for the vierbein and torsion
\begin{eqnarray}
e_\mu^i&=&{\rm Tr} \left[T^{5i}U_\mu\right]\\
T_{\mu\nu}^i&=&{\rm Tr} \left[T^{5i}\frac{1}{2}\left(U_{\mu\nu}-U_{\nu\mu}\right)\right]\;\;i=1,\ldots 4
\end{eqnarray}
while the $SO(4)$ components of the curvature 
${\cal R}_{\mu\nu}=R_{\mu\nu}-e_{\left[\mu\right.}e_{\left.\nu\right]}$
are given by
\beq
{\cal R}_{\mu\nu}^{ij}={\rm Tr}\left[T^{ij}\frac{1}{2}\left(U_{\mu\nu}-U_{\nu\mu}\right)\right]\;\;i=1\ldots 4\eeq

As in the previous studies \cite{Caracciolo:1988dg,Caracciolo:1987qi}
we have assumed an $SO(5)$ invariant Haar measure
on the group in the path integral defining the quantum theory. We have
employed a standard metropolis algorithm to perform the Monte
Carlo simulation. 

\section{Numerical results}
\subsection{Pure Palatini}

\begin{figure}
\begin{center}
\includegraphics[height=80mm]{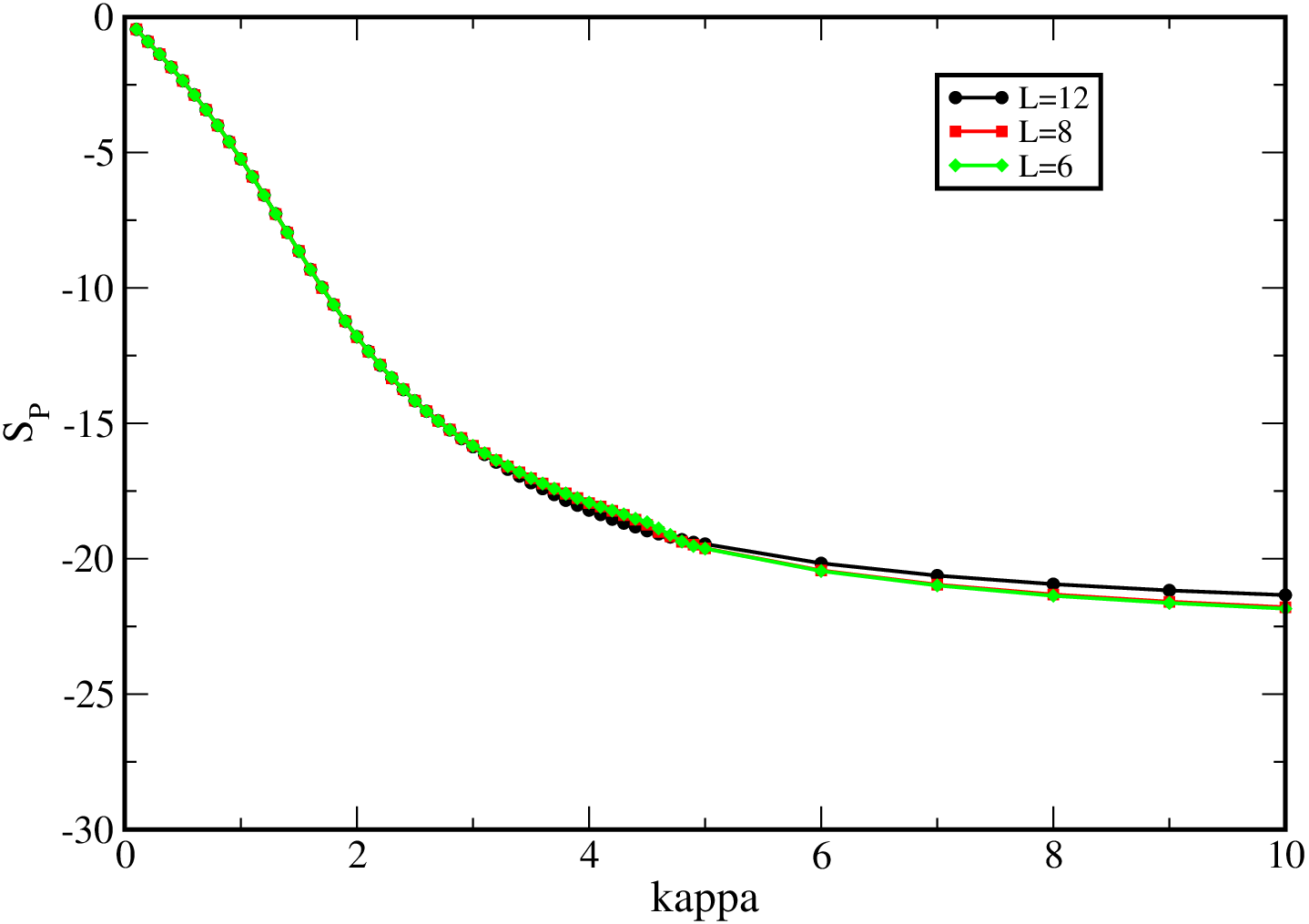}
\caption{Expectation value of Palatini action for $\alpha=0$ vs coupling $\kappa$}
\label{pal}
\end{center}
\end{figure}
\begin{figure}
\begin{center}
\includegraphics[height=80mm]{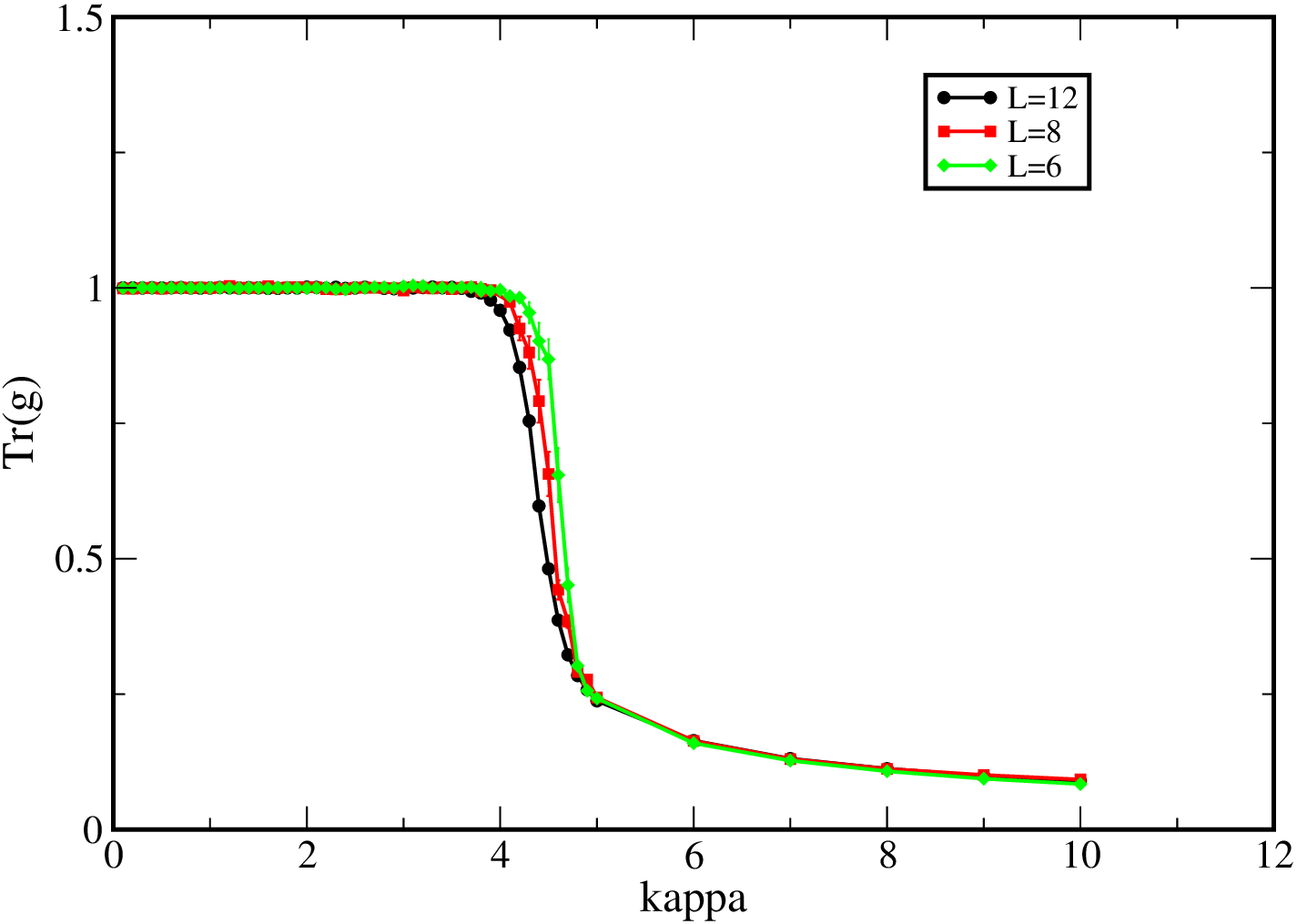}
\caption{Expectation value of ${\rm Tr}(g)$ for $\alpha=0$ vs coupling $\kappa$}
\label{viel}
\end{center}
\end{figure}

In fig.~\ref{pal} we show the expectation value of the pure Palatini
action ($\alpha=0$) as a function of the coupling $\kappa$ for
a sequence of three lattice sizes $L=6,8,12$. One sees
that the action approaches its lower kinematic bound of
$S_P=-24$ (which reflects the number of non-zeroes of the $\epsilon$ symbol)
independent of $L$ as the weak coupling limit $\kappa\to\infty$
is taken. This limit is reached when the plaquette variables
satisfy the extremality condition  
\beq
U_{\mu\nu}=-\gamma_5\epsilon_{\mu\nu\rho\lambda}U_{\rho\lambda}^\dagger
\label{extreme}\eeq
We will see later that pure Palatini
configurations satisfying this condition possess a maximal
$SO(4)$ curvature and hence cannot be interpreted as corresponding to
a smooth spacetime. Presumably this reflects
the usual unboundedness problem of Euclidean quantum
gravity. Furthermore, we observe that
the expectation value of the
vierbein approaches zero as $\kappa\to\infty$
as can be seen in fig.~\ref{viel} in which ${\rm Tr}(g)=\sum_{\mu, i} e_\mu^ie_\mu^i$ is plotted as
a function of $\kappa$ for the same range of lattice sizes\footnote{
We absorb the scale $l$ into our lattice vierbein throughout this
paper}. 
As emphasized earlier the vanishing of
the vierbein implies that the corresponding metric $g_{\mu\nu}$ is zero
and removes the possibility of interpreting the classical equations
of motion as corresponding to the field equations of a metric theory
of gravity.
 
\begin{figure}
\begin{center}
\includegraphics[height=80mm]{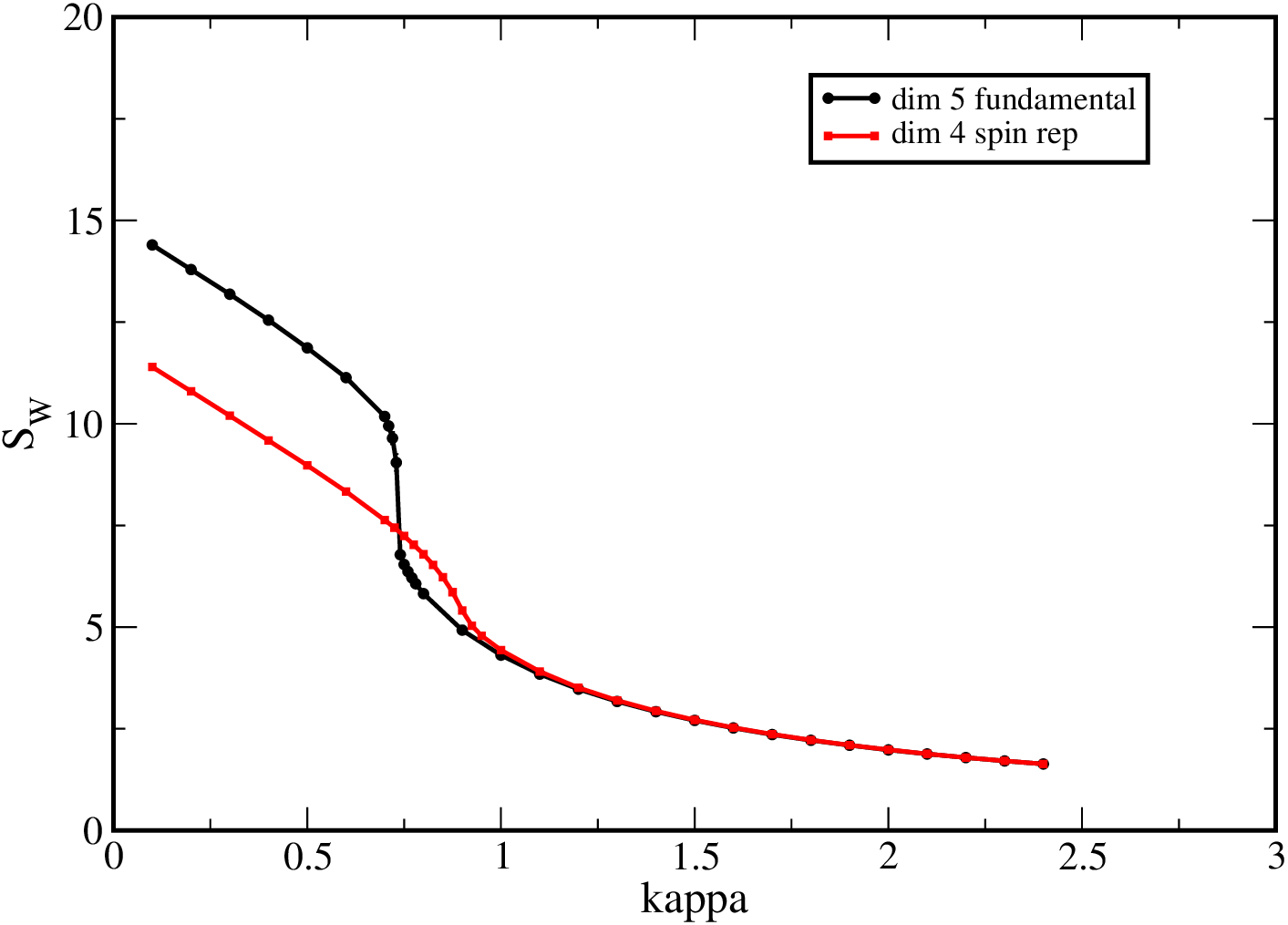}
\caption{Comparison of vector and spinor representations for $SO(5)$
Wilson action}
\label{compare}
\end{center}
\end{figure}

Before discussing the situation with $\alpha>0$ we first
return to an issue of what representation should be used for implementing
the local Lorentz invariance. As we have described in 
the last section, all the results shown in this paper have been
obtained 
using the four dimensional representation of spin(5) given in
terms of commutators of four dimensional Dirac matrices. 
As one can see in the previous two plots this choice generates
a smooth dependence of observables on the coupling $\kappa$. This is
quite different to what was seen in the only previous numerical study
of this system where a very strong first order phase transition was
observed separating strong from weak coupling.\footnote{One needs
to rescale $\kappa$ in this work by 16 to compare the two
couplings in these simulations \cite{Caracciolo:1988dg,Caracciolo:1987qi}} 
Indeed the observed hysteresis
effects were so severe in the latter case that it was hard to even thermalize 
the lattices
in the weak coupling phase. This earlier
study utilized the fundamental or vector representation
of $SO(5)$. 
This observed disparity in the phase structure of these two
lattice theories depending on choice of
representation is seen even in the pure Wilson theory ($\alpha=1$)
as can be seen in fig.~\ref{compare} which shows the average action
plotted as a function of $\kappa$ for the two representations
While the two representations agree at weak coupling as they
must since they possess the same
Lie algebra, they differ at strong coupling 
and the lattice action employing the fundamental representation
of $SO(5)$ suffers a strong first order transition for $\kappa\sim 0.75$ which
is completely absent in the spin(5) representation. This is analogous to
the situation for the lattice theories of $SO(3)$ and $SU(2)$ -- the former
exhibiting a first order bulk
phase transition separating weak from
strong coupling while the latter does not. Indeed, the analogy is
even stronger since $SU(2)\equiv spin(3)$ which leads us 
to conjecture that while
the $SO(N)$ groups have first order bulk transitions their covering groups
$spin(N)$ do not. Certainly from a practical point of view the
use of the spinor representation is clearly superior to that
of the vector representation.

\subsection{Palatini plus Wilson}
\begin{figure}
\begin{center}
\vspace{5mm}
\includegraphics[height=80mm]{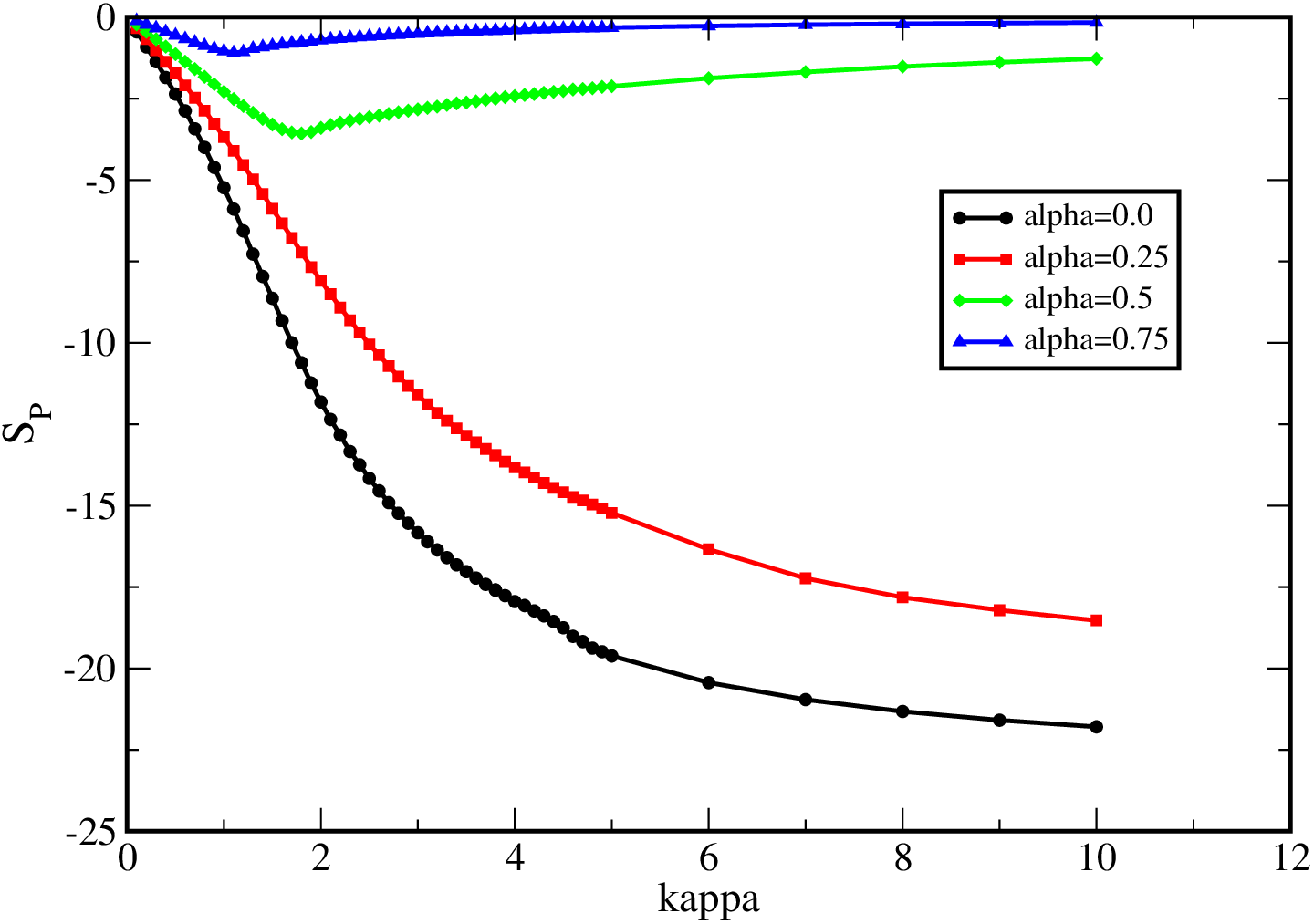}
\caption{Expectation value of Palatini action vs coupling $\kappa$ for various $\alpha$}
\label{palatini}
\end{center}
\end{figure}

\begin{figure}
\begin{center}
\vspace{10mm}
\includegraphics[height=80mm]{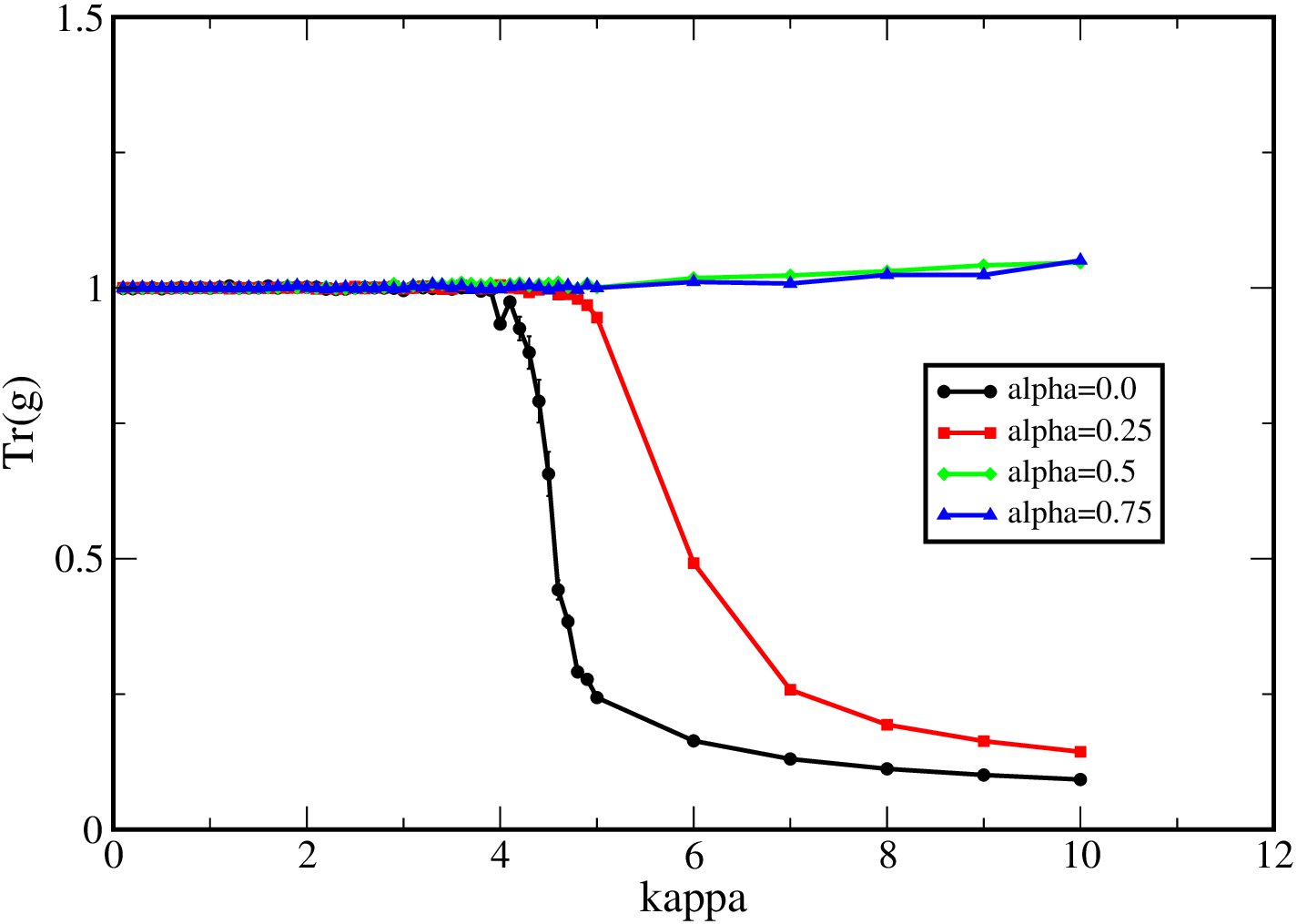}
\caption{Expectation value of ${\rm Tr}(g)$ vs coupling $\kappa$ for various $\alpha$}
\label{vierbein}
\end{center}
\end{figure}

\begin{figure}
\begin{center}
\includegraphics[height=80mm]{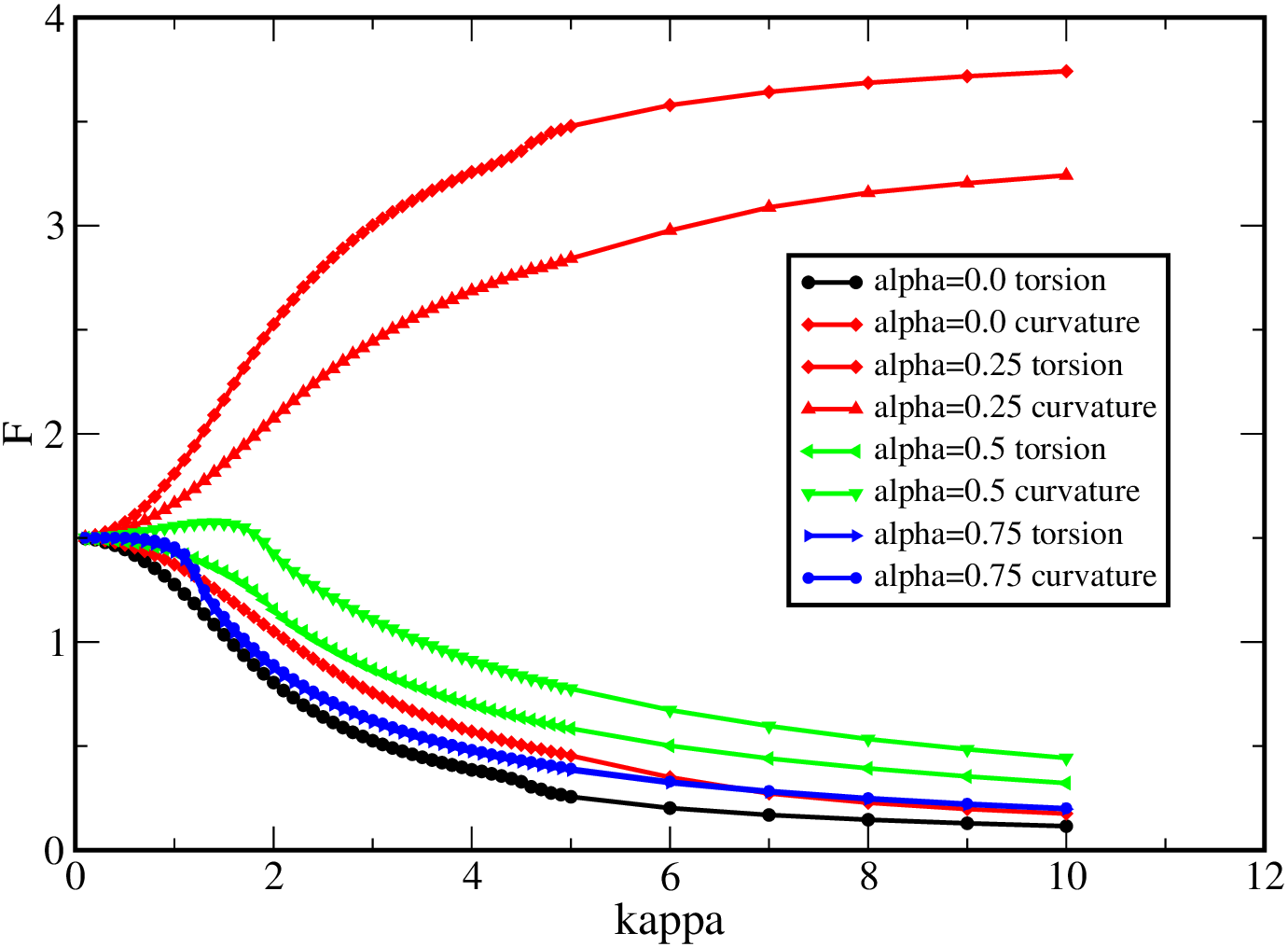}
\caption{Curvature and torsion components of the $SO(5)$ field strength vs $\kappa$}
\label{field}
\end{center}
\end{figure}

In this section we shows results obtained in the model with $\alpha>0$.
First consider the expectation value of the Palatini action in such models. 
Fig.~\ref{palatini} shows this for four different values of the coupling
$\alpha=0.0,0.25,0.5,0.75$ for a fixed $8^4$ lattice as a function
of $\kappa$. Clearly the curves fall into 2 classes; for small $\alpha$
the expectation value of the
action approaches the kinematic boundary corresponding to maximal
(negative) curvature while for larger $\alpha$ they approach  
vanishing action in the weak coupling limit. A similar contrasting
behavior is
seen in the plot of the vierbein as revealed in fig.~\ref{vierbein}.
At small $\alpha$ the vierbein is driven to zero for large $\kappa$ as for
pure Palatini while
for $\alpha$ above some critical value $\alpha_T$ the vierbein remains non-zero in
the weak coupling limit. Thus it appears that a sufficiently large
coupling to the Wilson operator stabilizes the vierbein. However
this is not enough -- we also require a zero torsion condition
and a small curvature. 
We now turn to these other observables.
Fig.~\ref{field} shows a plot of 
${\cal R}_{\mu\nu}^{ij}{\cal R}_{\mu\nu}^{ij}$ and $T_{\mu\nu}^iT_{\mu\nu}^i$
for the same values of $\alpha$ as a function of the coupling $\kappa$.
For small $\alpha$ the expectation values are driven to values similar
to the case for pure Palatini -- the $SO(4)$ curvature attaining
near maximal (negative) values as $\kappa\to\infty$\footnote{
Notice that ${\cal R}\to R$ for $e=0$}. However, again,
for large values of $\alpha$ both the
torsion $T$ {\it and} curvature ${\cal R}$ approach zero at weak coupling.
Indeed, notice that for $\alpha=0.75$ the torsion and curvature are degenerate to
within small errors in this limit.

\begin{figure}
\begin{center}
\vspace{5mm}
\includegraphics[height=80mm]{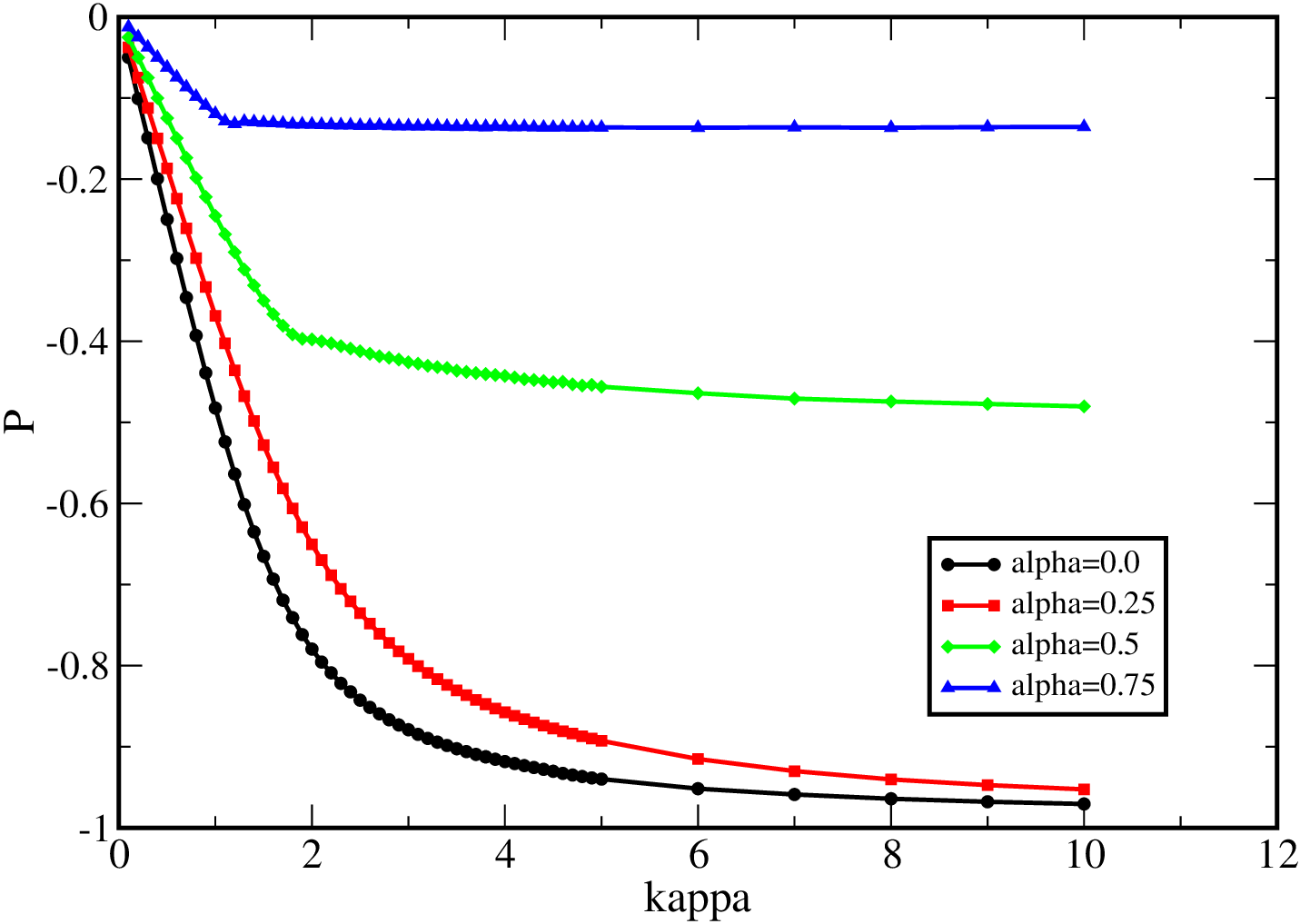}
\caption{Order parameter $P$ as a function of $\kappa$ for several $\alpha$}
\label{self}
\end{center}
\end{figure}

These are quite encouraging results but one might worry that the Palatini
action is playing no role at all for large $\alpha$ and the physics is
being dominated by just the Wilson term -- which would invalidate any
connection to gravity. To show that this is not the case we have
plotted a ``self-dual'' order parameter given by\footnote{This is
analogous to the self-dual order parameter introduuced in 
\cite{Caracciolo:1987qi} although
in that case the duality operation involves only the internal space}
\beq
P=\frac{\sum_x \epsilon_{\mu\nu\lambda\rho}\epsilon^{ijkl}
U_{\mu\nu}^{ij}U_{\lambda\rho}^{kl}}{\sum_x U_{\mu\nu}^{ij}U_{\mu\nu}^{ij}}\eeq
On the extremal configurations given by eqn.~\ref{extreme}
this attains a value of minus one. This should thus be the case for
pure Palatini at large $\kappa$. In contrast it should vanish in the case
of the pure Wilson action.
In fig.~\ref{self} we show the value of this order parameter for the same
range of $\alpha$ and $\kappa$ on the $8^4$ lattice.
Notice that the curves for $\alpha=0.5,0.75$ indicate that 
order parameter appears to approach a constant for large
$\kappa$ at a value which  is intermediate between the pure Palatini and
Wilson values. We regard this as a piece of evidence disfavoring a system
governed solely by the Wilson term.

\begin{figure}
\begin{center}
\includegraphics[height=80mm]{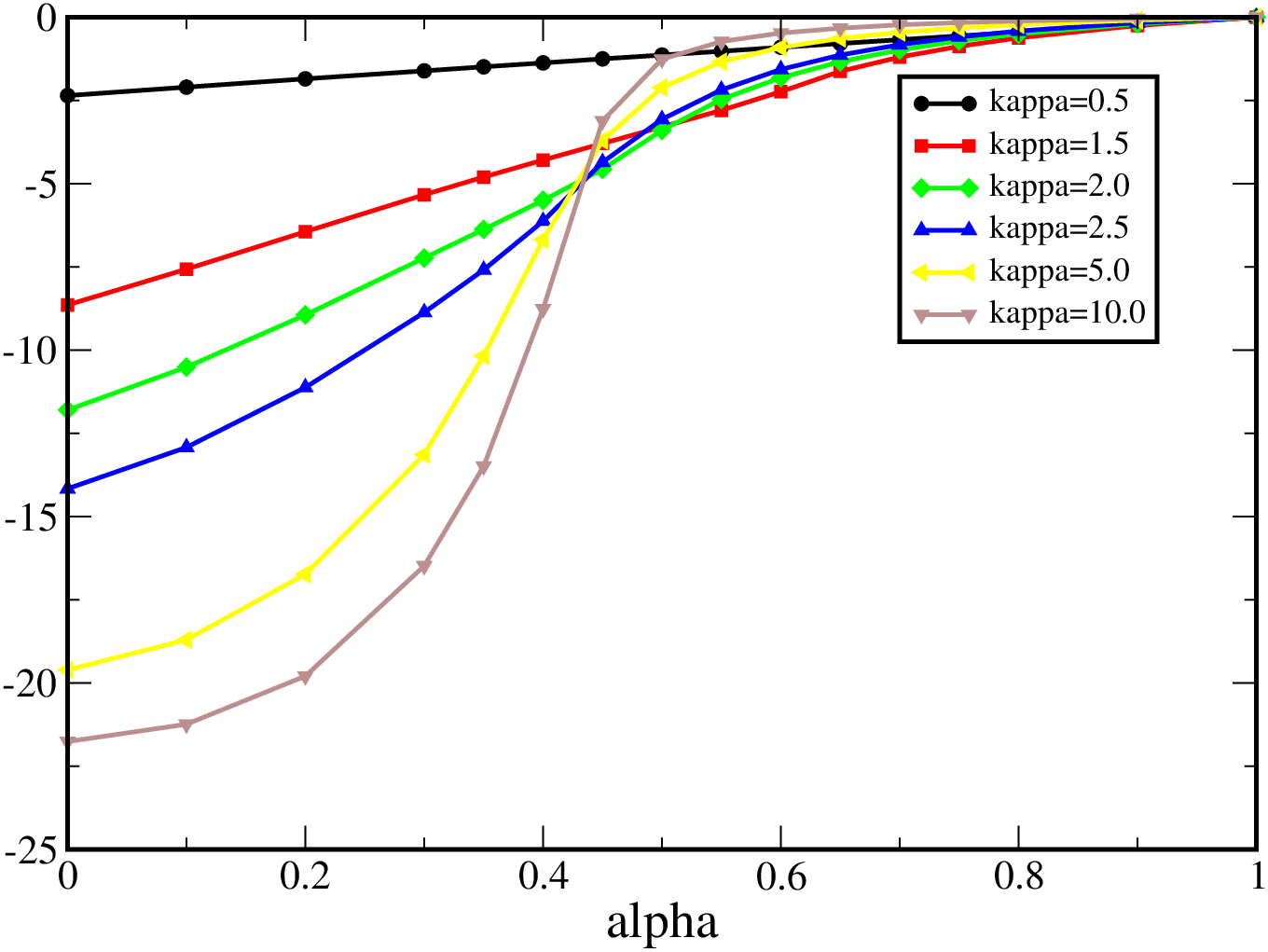}
\caption{Expectation value of Palatini action as function of $\alpha$ for several $\kappa$}
\label{newpal}
\end{center}
\end{figure}

Up to this point we have concentrated
on showing results as a function of $\kappa$ for several
discrete values of $\alpha$.
However it is also
very instructive to show the data as function of $\alpha$ at fixed
$\kappa$. For example, the expectation value of the Palatini action is shown
in fig.~\ref{newpal}
For $\alpha<\alpha_T$ with $\alpha_T=0.4-0.5$ it is clear that the
Palatini action flows to its kinematic lower limit as $\kappa\to\infty$
independent of the value of $\alpha$ within this range
while for $\alpha>\alpha_T$ the trend is reversed and the action goes to
zero for large enough $\kappa$. This suggests two different
ground states are available to the system at least for large
enough $\kappa$ whose domain of attraction is determined
by the coupling $\alpha$. For $\alpha<\alpha_T$ the system
at weak coupling is governed by
the pure Palatini action while for $\alpha>\alpha_T$ the system
flows to a ground state corresponding to vanishing $SO(5)$
curvature. In the limit $\alpha\to 1$ this is just the usual
Wilson vacuum. To contrast with this behavior notice that for small
$\kappa$ coupling eg $\kappa=0.5$, the value of 
the action depends only very weakly
on the associated coupling $\alpha$ and a system seems to
exist in just one phase. Of course one expects that the
ground state of the system for small $\kappa$ is dominated by
lattice artifacts so this latter
phase is ultimately uninteresting. 

\begin{figure}
\begin{center}
\vspace{5mm}
\includegraphics[height=80mm]{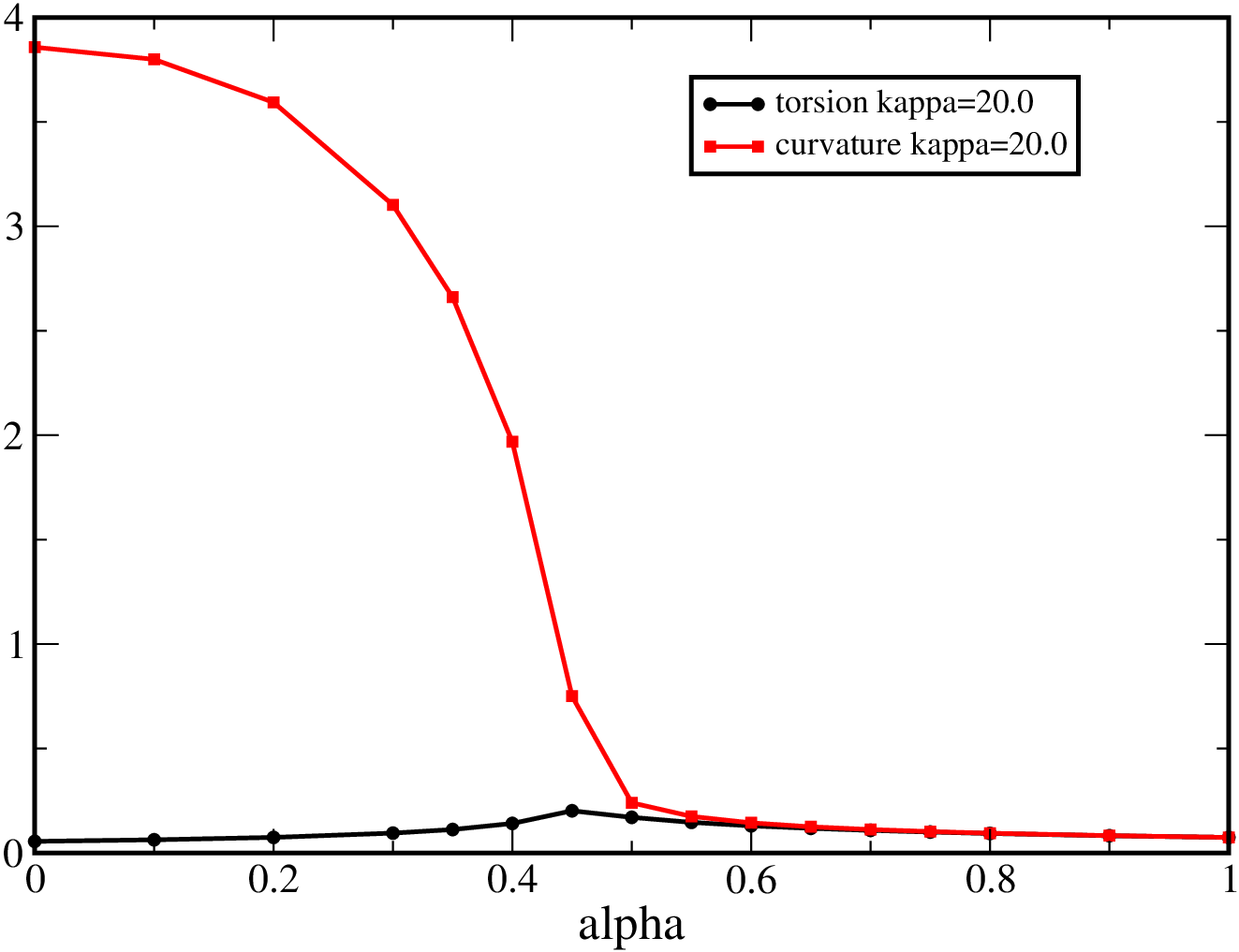}
\caption{Curvature and torsion as a function of $\alpha$ for $\kappa=20.0$}
\label{tor}
\end{center}
\end{figure}

This cross-over between two different
ground states for large $\kappa$
is clearly seen by looking at the torsion and
curvature as a function of $\alpha$.
A plot (fig.~\ref{tor}) showing $T$ and ${\cal R}$ for
$\kappa=20.0$ as a function
of $\alpha$ confirms the importance of this threshold coupling $\alpha_T$
As $\alpha$ increases the curvature drops towards to zero rapidly merging
with the torsion for $\alpha>\alpha_T\sim 0.5$.

\begin{figure}
\begin{center}
\vspace{5mm}
\includegraphics[height=80mm]{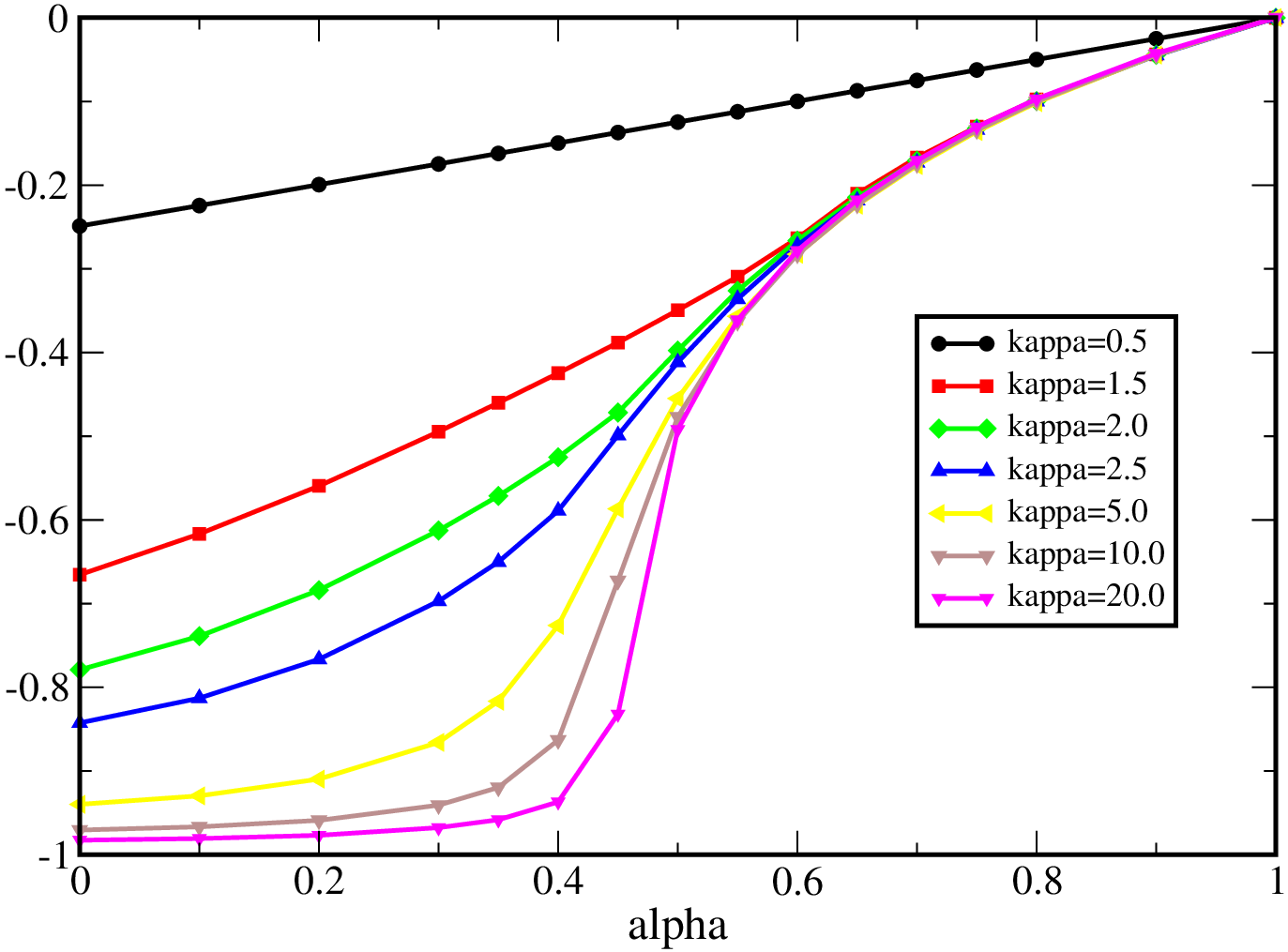}
\caption{Order parameter as a function of $\alpha$ for several $\kappa$}
\label{newself}
\end{center}
\end{figure}

A similar picture is revealed by looking at the plot of the order parameter $P$
as shown in fig.~\ref{newself}. 
It appears that the order parameter equals minus one for all $\alpha<\alpha_T$
as $\kappa\to\infty$ but then rises towards zero as $\alpha$ varies in the
range $\alpha>\alpha_T$. 
Notice also
that there
is little $\kappa$ dependence in the curves for $\alpha>\alpha_T$ - the
curves for different $\kappa$ rapidly approach fixed well-defined
envelope in this region of parameter space.
Again, let us emphasize that the small $\kappa$ data doesn't fit this picture -- there is
no evidence for a threshold value of $\alpha$ in the $\kappa=0.5$ curve. 
Thus the phase diagram seems to contain a single strong coupling phase but the
possibility of two dramatically different phases at large $\kappa$ coupling.
It is unclear from the data whether there
is a discontinuity or not for $\alpha\to\alpha_T$. One hint favoring
a true phase transition at $\alpha_T$ can be obtained by
looking at the local volume element 
$\epsilon^{ijkl}\epsilon_{\mu\nu\lambda\rho}e^i_\mu e^j_\nu e^k_\lambda e^l_\rho$.
We can extract a gauge invariant measure for this on the lattice
by extracting an expression for a matrix valued representation
of the vierbein at a site:
\beq
e_\mu(x)=\gamma_5 U_\mu(x)\gamma_5 U^\dagger_\mu(x)-I
\eeq
which leads to an expression for the gauge invariant volume element
\beq
\sqrt{g}=\epsilon_{\mu\nu\lambda\rho}{\rm Tr}
\left(e_\mu e_\nu e_\lambda e_\rho\right)\eeq
\begin{figure}
\begin{center}
\vspace{5mm}
\includegraphics[height=80mm]{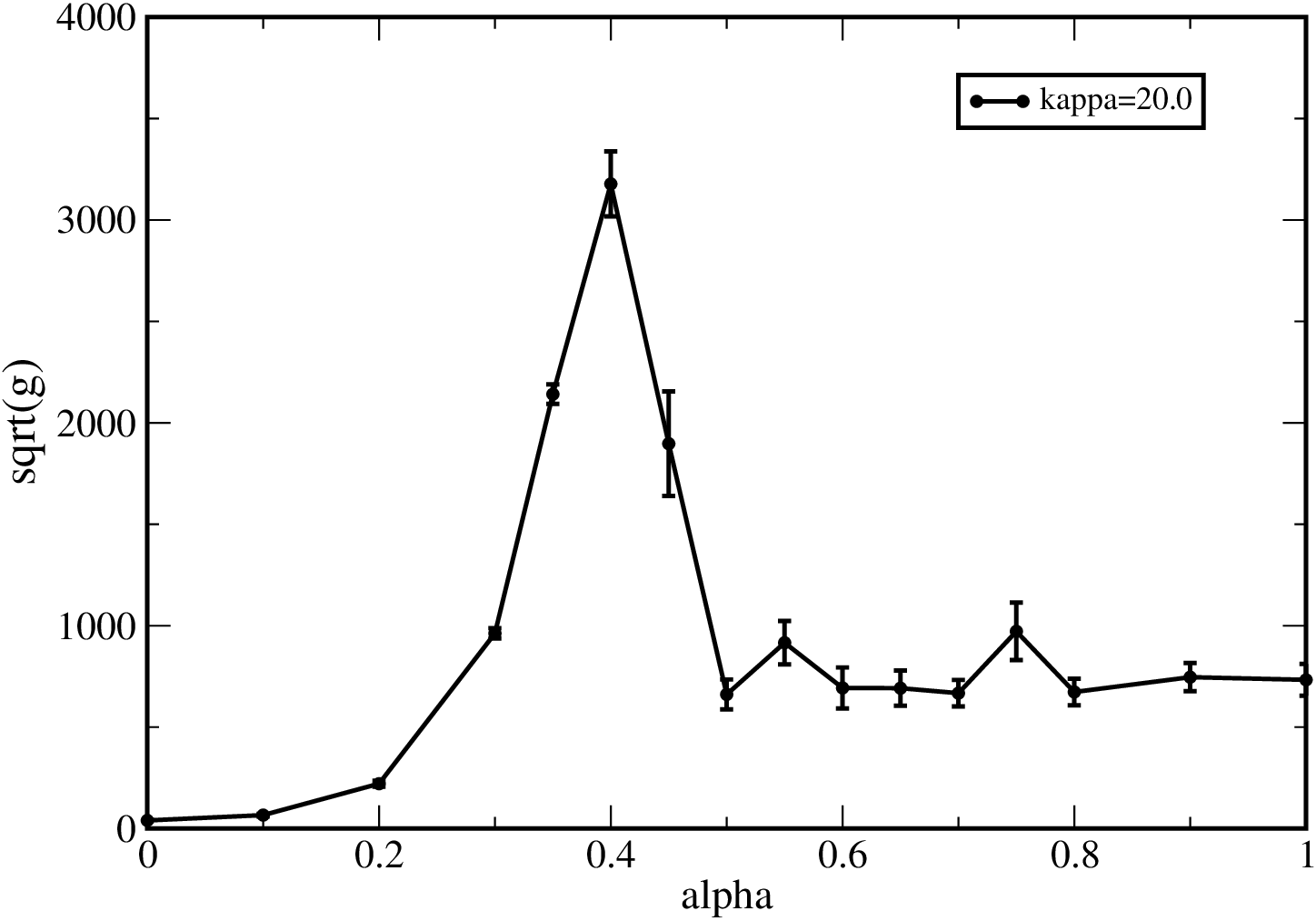}
\caption{Absolute value of the volume element vs $\alpha$ for $\kappa=20.0$}
\label{newvol}
\end{center}
\end{figure}
The absolute value of this quantity is shown in fig.~\ref{newvol} for 
$\kappa=20.0$ and $L=8$
as a function of $\alpha$. For small $\alpha$ the vierbein vanishes and
so does the volume element while for $\alpha>\alpha_T$ it becomes
constant. However, in the vicinity of
$\alpha_T$ we see a sharp spike corresponding
to a large local volume element in lattice units.
\section{Discussion}

We have simulated a Euclidean model for quantum gravity based on
discretization of the Palatini action including a cosmological
constant term and supplemented by a Wilson term. Unlike a previous study of the
pure Palatini action we have employed the spinor representation of the
spin(5) covering group to define the lattice action. This is faster
to implement and appears to avoid some of the strong lattice artifacts
seen in earlier studies of this action. 

We show that the pure Palatini lattice theory is sick suffering from
a local curvature which attains the maximal (negative) value consistent with the
compact gauge symmetry, a vanishing vierbein and lattice doublers.

We argue that the Wilson operator serves both to regulate all of these
problems and furthermore will necessarily be induced at the
quantum level if it is not present in the bare lattice action. In the
expanded parameter space determined by the Palatini and
Wilson couplings we find evidence for two new ground states for
sufficiently weak coupling (large $\kappa$) -- one
corresponding to pure Palatini and suffering from
all the old problems and a second in which the vierbein is non-zero,
and the torsion and curvature are small. These two vacua can be
distinguished by a self-dual order parameter and can be accessed
by tuning the coupling $\alpha$ which controls the mixing with
the Wilson term. 

It appears that for $\alpha$
couplings close to some threshold value $\alpha_T$ the new vacuum is {\it not} just
that of the pure Wilson action but retains a memory of the
Palatini term as revealed by the self-dual order parameter acquiring a
non-trivial value there. It remains to be seen whether this threshold value
$\alpha_T$ can be thought of as a true critical value 
and if so whether
this critical value corresponds to a continuous or discontinuous phase
transition. The latter issue is of course a crucial issue to address
in the context of obtaining a non-trivial continuum limit.

\acknowledgments SMC and DF are supported in part by DOE grant
DE-FG02-85ER40237. The simulations were carried out using USQCD
resources at Fermilab.

%
\bibliographystyle{JHEP}
\bibliography{dS}
%

\end{document}